\documentclass[sn-mathphys,Numbered]{sn-jnl}

\usepackage{graphicx}%
\usepackage{multirow}%
\usepackage{amsmath,amssymb,amsfonts}%
\usepackage{amsthm}%
\usepackage{mathrsfs}%
\usepackage[title]{appendix}%
\usepackage{xcolor}%
\usepackage{textcomp}%
\usepackage{manyfoot}%
\usepackage{booktabs}%
\usepackage{algorithm}%
\usepackage{algorithmicx}%
\usepackage{algpseudocode}%
\usepackage{listings}%
\usepackage{verbatim}%

\theoremstyle{thmstyleone}%
\theoremstyle{thmstyletwo}%

\theoremstyle{thmstylethree}%

\begin{document}

\title{Direct Experimental Constraints on the Spatial Extent of a Neutrino Wavepacket}

\author*[1]{\fnm{Joseph} \sur{Smolsky}}\email{joseph.smolsky@mines.edu}
\author*[1,2]{\fnm{Kyle} G \sur{Leach}}\email{kleach@mines.edu}
\author[3]{\fnm{Ryan} \sur{Abells}}
\author[4]{\fnm{Pedro} \sur{Amaro}}
\author[5]{\fnm{Adrien} \sur{Andoche}}
\author[1]{\fnm{Keith} \sur{Borbridge}}
\author[1,6]{\fnm{Connor} \sur{Bray}}
\author[7]{\fnm{Robin} \sur{Cantor}}
\author[8]{\fnm{David} \sur{Diercks}}
\author[1]{\fnm{Spencer} \sur{Fretwell}}
\author[6]{\fnm{Stephan} \sur{Friedrich}}
\author[1]{\fnm{Abigail} \sur{Gillespie}}
\author[4]{\fnm{Mauro} \sur{Guerra}}
\author[7]{\fnm{Ad} \sur{Hall}}
\author[1]{\fnm{Cameron} N \sur{Harris}}
\author[9]{\fnm{Jackson} T \sur{Harris}}
\author[1]{\fnm{Calvin} \sur{Hinkle}}
\author[1]{\fnm{Amii} \sur{Lamm}}
\author[10]{\fnm{Leendert} M \sur{Hayen}}
\author[5]{\fnm{Paul-Antoine} \sur{Hervieux}}
\author[6]{\fnm{Geon-Bo} \sur{Kim}}
\author[6]{\fnm{Inwook} \sur{Kim}}
\author[3,11]{\fnm{Annika} \sur{Lennarz}}
\author[6]{\fnm{Vincenzo} \sur{Lordi}}
\author[4]{\fnm{Jorge} \sur{Machado}}
\author[1]{\fnm{Andrew} \sur{Marino}}
\author[3]{\fnm{David} \sur{McKeen}}
\author[12]{\fnm{Xavier} \sur{Mougeot}}
\author[13]{\fnm{Francisco} \sur{Ponce}}
\author[3]{\fnm{Chris} \sur{Ruiz}}
\author[6]{\fnm{Amit} \sur{Samanta}}
\author[4]{\fnm{Jos\'e Paulo} \sur{Santos}}
\author[1]{\fnm{Caitlyn} \sur{Stone-Whitehead}}
\author[1]{\fnm{John} \sur{Taylor}}
\author[1]{\fnm{Joseph} \sur{Templet}}
\author[3]{\fnm{Sriteja} \sur{Upadhyayula}}
\author[3]{\fnm{Louis} \sur{Wagner}}
\author[9]{\fnm{William} K \sur{Warburton}}

\affil[1]{\orgdiv{Department of Physics}, \orgname{Colorado School of Mines}, \orgaddress{\street{1500 Illinois St}, \city{Golden}, \postcode{80401}, \state{Colorado}, \country{USA}}}

\affil[2]{\orgdiv{Facility for Rare Isotope Beams}, \orgname{Michigan State University}, \orgaddress{\street{640 S Shaw Lane}, \city{East Lansing}, \postcode{48824}, \state{MI}, \country{USA}}}

\affil[3]{TRIUMF, 4004 Wesbrook Mall, Vancouver, BC V6T 2A3, Canada}
\affil[4]{LIBPhys-UNL, Departamento de Física, Faculdade de Ciências e Tecnologia, NOVA FCT, Universidade Nova de Lisboa, 2829-516 Caparica, Portugal}
\affil[5]{Universit\'e de Strasbourg, CNRS, Institut de Physique et Chimie des Mat\'eriaux de Strasbourg,UMR 7504, F-67000 Strasbourg, France}
\affil[6] {Lawrence Livermore National Laboratory, Livermore, CA 94550, USA}
\affil[7]{STAR Cryoelectonics LLC, Santa Fe, NM 87508, USA}
\affil[8]{Shared Instrumentation Facility, Colorado School of Mines, Golden, CO 80401, USA}
\affil[9]{XIA LLC, Hayward, CA 94544, USA}
\affil[10]{LPC Caen, ENSICAEN, Université de Caen, CNRS/IN2P3, Caen, France}
\affil[11]{Department of Physics and Astronomy, McMaster University, Hamilton, Ontario L8S 4M1, Canada}
\affil[12]{Universit\'e Paris-Saclay, CEA, List, Laboratoire National Henri Becquerel (LNE-LNHB), F-91120, Palaiseau, France}
\affil[13]{Pacific Northwest National Laboratory, Richland, WA 99354, USA}

\abstract{Despite their high relative abundance in our Universe, neutrinos are the least understood fundamental particles of nature.  They also provide a unique system to study quantum coherence and the wavelike nature of particles in fundamental systems due to their extremely weak interaction probabilities.  In fact, the quantum properties of neutrinos emitted in experimentally relevant sources are virtually unknown and the spatial extent of the neutrino wavepacket is only loosely constrained by reactor neutrino oscillation data with a spread of 13 orders of magnitude.  Here, we present the first direct limits of this quantity through a new experimental concept to extract the energy width, $\sigma_{\textrm{N},E}$, of the recoil daughter nucleus emitted in the nuclear electron capture (EC) decay of $^7$Be.  The final state in the EC decay process contains a recoiling $^7$Li nucleus and an electron neutrino ($\nu_e$) which are entangled at their creation.  The $^7$Li energy spectrum is measured to high precision by directly embedding $^7$Be radioisotopes into a high resolution superconducting tunnel junction that is operated as a cryogenic sensor.  The lower limit on the spatial uncertainty of the recoil daughter was found to be $\sigma_{\textrm{N}, x} \geq 6.2$\,pm, which implies the final-state system is localized at a scale more than a thousand times larger than the nucleus itself.  From this measurement, the first direct lower limits on the spatial extent of the neutrino wavepacket were extracted using two different theoretical methods.  These results have wide-reaching implications in several areas including the nature of spatial localization at sub-atomic scales, interpretation of neutrino physics data, and the potential reach of future large-scale experiments.}


\maketitle

Quantum mechanics is based on the concept that the microscopic scale of our Universe contains inherent uncertainties in position and momentum that are fundamentally connected via $\sigma_x \sigma_p \geq \hbar/2$~\cite{Heisenberg1925}.  Using this theoretical description, precise knowledge of the momentum width of the particle, $\sigma_p$, implies a corresponding uncertainty on the extent of its spatial wavefunction, $\sigma_x$.  When the object loses its ability to maintain coherence through interactions with the complex environmental bath, its position is localized.  This spatial localization of quantum objects depends heavily on the specific environment they are subjected to~\cite{RevModPhys.91.021001}.

Recent efforts have primarily focused on constructing increasingly large coherent quantum systems~\cite{Fein2019, doi:10.1126/science.adf7553} and other highly delocalized objects to push measurement and other applications to, and beyond, the standard quantum limit~\cite{beckey2023quantum,Bass2024}.  Such experiments are at the frontiers of quantum science and engineering and aim, in part, to probe the interface between classical and quantum mechanics.  Here, we present a new measurement concept on the other side of the coin; using highly localized unstable systems embedded in a complex solid-state material at low temperature as a precision laboratory to probe quantum properties of subatomic particles.  We show that this concept is particularly powerful for investigating properties of systems that are otherwise not directly accessible due to weak couplings in the Standard Model of particle physics (SM), in particular the neutrino ($\nu$).

Neutrinos are light, neutral leptons, and the only particles in the SM that have an intrinsic chirality, in that they only interact via \textit{left-handed} currents of the weak interaction through three eigenstates, $\nu_e$, $\nu_\mu$, and $\nu_\tau$.  Neutrino oscillation experiments over the last 30 years~\cite{Fuk98,Ahm01} have also indicated that these flavour states include at least two non-zero mass eigenstates.  This observation of non-zero neutrino masses currently provides the only confirmed violation of the SM as it was originally constructed~\cite{Ramond:1999vh,Bilenky:2014ema}, for which the Nobel Prize in physics was awarded in 2015~\cite{Taroni2015}.  The flavour oscillation effect for neutrinos arises from the fact that the mass eigenstates and weak-interaction eigenstates are not equal to each other, and due to their small interaction probabilities with the weak and gravitational forces, are able to maintain coherence over long distances.  Thus, even for a neutrino created with a definite flavour -- the electron neutrino ($\nu_e$) in weak nuclear decay, for example -- a coherent superposition of states exists within the neutrino wavepacket that affects the probability for observing a particular flavour as a function of time or distance from the source.

Since neutrinos only interact weakly with matter, they are notoriously difficult to detect\footnote{For example, a $\nu_e$ with 1~MeV kinetic energy has an interaction cross section of roughly $10^{-42}$~cm$^{2}$.}~\cite{RevModPhys.84.1307}.  This fact also makes neutrinos attractive systems to study fundamental properties of quantum mechanics and exotic new physics~\cite{Abbasi2024}, since they are able to maintain coherence over long distances.  Historically, observing these so-called ``ghost particles" from various sources requires large volume detectors with limited detection precision that run for several years~\cite{NuDetectorReview2023}.  These detectors rely on neutral current (NC) and charged current (CC) interactions with an atomic nucleus or electron within these large volumes~\cite{RevModPhys.84.1307}.  The final state particles are used to reconstruct the incoming neutrino energy and interaction point, along with neutrino flavour for CC interactions.  A deficit or excess flux of any flavour in the detector, compared to the flavour ratios at the source, is used to determine the oscillation probabilities as a function of energy and distance.   Global data from neutrino oscillation experiments have been used to determine the mass-splittings and mixing parameters for the three SM neutrinos~\cite{PDG2023}.  Data from the Daya Bay~\cite{CAO201662,PhysRevLett.129.041801}, RENO~\cite{renocollaboration2010reno, PhysRevLett.108.191802}, and KamLAND~\cite{the_kamland_collaboration_constraints_2011} experiments currently also provide loose constraints on the spatial extent of $\overline{\nu}_e$ wavepackets from $\beta^-$ decay (reactor sources) of $2.1\times10^{-13}$~m~$ \leq \sigma_{\nu,x} \leq 2$\,m~\cite{daya_bay_collaboration_study_2017,de_gouvea_probing_2020,de_gouvea_combined_2021}.  If the $\overline{\nu}_e$ and $\nu_e$ wavepacket spatial widths are indeed near the low end of these limits, it has a significant impact in the neutrino physics landscape, including alleviating tension in models for sterile neutrinos with eV-scale masses through model-dependent fits to available data~\cite{arguelles_impact_2023, hardin_new_2023}.

To extract physics from the oscillation data, however, interpretation within the lepton mixing framework of the SM is required~\cite{PDG2023}.  Given the current pervading anomalies in the neutrino physics landscape~\cite{universe7100360}, there is a need for new experimental techniques that can provide model-independent access to the neutrino.  Clever new experimental concepts have been presented to address this need using efficient, small scale setups~\cite{Leach2022,Martoff2022,PRXQuantum.4.010315} that are less model-dependent and not limited by low neutrino interaction cross sections.  These experiments aim to precisely measure the low-energy recoiling atoms in nuclear electron capture (EC) decay~\cite{RevModPhys.49.77} to access information of the neutrino directly through its entanglement in the final state of radioactive decay.  Of these new concepts, the beryllium electron capture in superconducting tunnel junctions (BeEST) experiment~\cite{Leach2022} is currently the only one to practically employ the concept of direct EC daughter recoil detection for neutrino physics~\cite{PhysRevLett.126.021803} and astrophysics~\cite{Fre20}.  In this paper, we report the first direct limit on the spatial width of a neutrino wavepacket from the energy width of the recoiling nucleus in EC decay, $\sigma_{\textrm{N}, E}$, through their mutual entanglement.

Nuclear EC decay is a radioactive decay mode that results from the capture of an $s$-wave orbital electron by a proton within the nuclear volume~\cite{RevModPhys.49.77}.  At the fundamental level, this weak interaction process transmutes an up quark to a down quark through the exchange of a $W$ boson within the nucleon and conserves lepton number through the emission of $\nu_e$.  Within the experimental context of the work presented here, the observable final state only contains two products: the recoiling heavy daughter system and $\nu_e$.  This simple two-body system that we observe is entangled at its creation, and thus a precision measurement of the daughter recoil provides direct access to information on the neutrino.

With seven nucleons and four electrons, the neutron deficient nucleus beryllium-7 ($^7$Be) is the simplest pure EC decaying system in the nuclear landscape.  $^7$Be decays to lithium-7 ($^7$Li) and $\nu_e$ with a half-life of $T_{1/2}=53.22(6)$~days~\cite{Til02} and a total decay energy of $Q_{EC}=861.963(23)$~keV~\cite{PhysRevC.109.L022501}.  A small branch of 10.44(4)\% results in the population of a short-lived excited nuclear state in $^7$Li ($T_{1/2}=72.8(20)$~fs)~\cite{Til02} that de-excites via emission of a 477.603(2)~keV $\gamma$-ray~\cite{Hel00}.  In the EC process, the electron can be captured either from the $1s$ shell ($K$-capture) or the $2s$ shell ($L$-capture) of Be.  For $K$-capture, the binding energy of the $1s$ hole is subsequently liberated by emission of an Auger electron~\cite{Hub94}.  Since the nuclear decay and subsequent atomic relaxation occur on short time scales, a direct measurement produces a spectrum with four peaks: two for $K$-capture and two for $L$-capture into the ground state (GS) and the excited state (ES) of $^7$Li.

\begin{figure}[t!]
\centering
\includegraphics[width=0.8\textwidth]{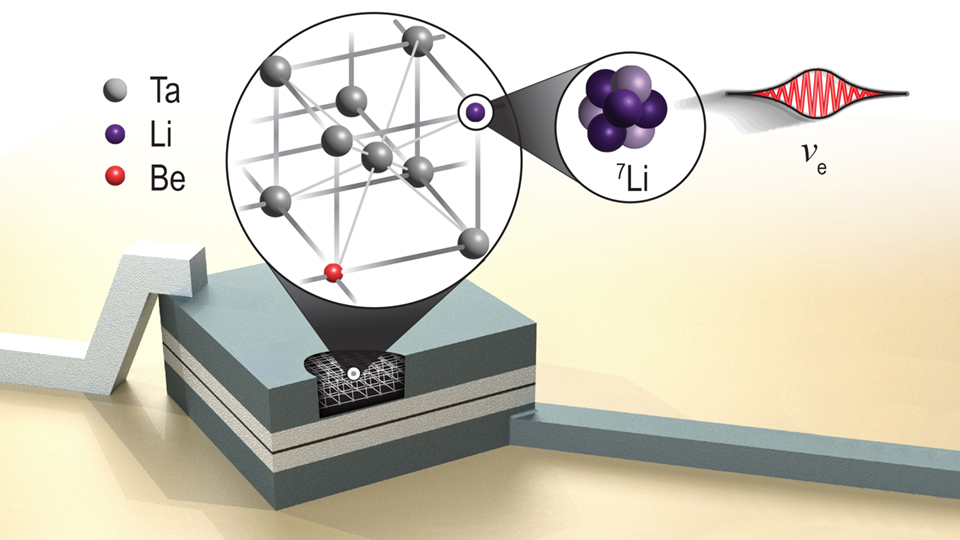}
\caption{\label{fig:schematic} A schematic image of the STJ sensor showing the five layers described in the text.  The $^7$Be radioactive source is sparsely implanted into the Ta lattice of the absorber layer.  The upper right zoom-in shows the final state after EC decay including the $^7$Li recoil and $\nu_e$.}
\end{figure}

To measure the $^7$Be EC decay spectrum to high precision, we directly embedded the $^7$Be into the top tantalum film of a superconducting tunnel junction (STJ) pixel that was used as a high energy-resolution sensor~\cite{Friedrich2008}.  The single STJ pixel used in this experiment was part of a 36-pixel array where each sensor had a surface area of $208\times208$~$\mu$m$^2$ and consisted of five layers (from top to bottom): Ta (165 nm)—Al (50 nm)—Al$_2$O$_3$ (1 nm)—Al (50 nm)—Ta (265 nm), shown in Fig.~\ref{fig:schematic}, that were fabricated by photolithography at STAR Cryoelectronics LLC~\cite{Car13}.  STJs exploit the small energy gap in superconducting Ta ($\Delta_{Ta}\approx0.7$~meV) to provide $\sim30\times$ higher energy resolution than conventional Si or Ge detectors, thus allowing for a precision measurement of the $^7$Li recoil energies.

Radioactive $^7$Be was implanted directly into the top tantalum film of the STJ pixel through apertures in a Si collimator that was situated $\sim100~\mu$m above the chip.  The $^7$Be was produced and delivered to the implantation chamber end-station at the TRIUMF-ISAC facility in Vancouver, Canada~\cite{Dil14} at an energy of 30~keV.  The isotopically pure $(\sim90\%)$ $^7$Be$^+$ beam was produced using the isotope separation on-line (ISOL) technique~\cite{Blu13} via spallation reactions from a 10~$\mu$A, 480-MeV proton beam incident on a stack of thin uranium carbide targets.  Once released from the target through diffusion, the created beryllium atoms were selectively ionized using the ion guide laser ion source (IG-LIS)~\cite{10.1063/1.4868496} operated in suppression mode~\cite{Mostamand2020}.  The resulting beam of $^7$Be$^+$ ions was implanted at a rate of $6.2\times10^6$ $s^{-1}$ for a period of 25 hours to generate an initial per-pixel activity in the STJs of roughly 50 Bq.  Following implantation, the chip was cleaned and processed at TRIUMF to remove loose and surface-deposited activity and subsequently shipped to Lawrence Livermore National Laboratory (LLNL).

The $^7$Li recoil spectrum from the decay of $^7$Be was measured at LLNL with the STJ detector at a temperature of $\sim$0.1 K in a two-stage adiabatic demagnetization refrigerator (ADR) with liquid nitrogen and liquid helium precooling. Signal traces were read out continuously at a rate of 1.25 MSa/s at a 16 bit resolution using a PXIe-6356 ADC.  For real-time \textit{in-situ} monitoring of the response and energy calibration, the STJs were simultaneously exposed to 3.49865(15)~eV photons from a pulsed, 355~nm frequency tripled Nd:YVO$_4$ laser triggered at a rate of 100~Hz that was fed into the cold stage of the ADR through a fiber to illuminate the pixels. The laser intensity was adjusted such that multi-photon absorption provided a comb of peaks over the energy range from $20-120$~eV.

\begin{figure}[t!]
\centering
\includegraphics[width=0.7\textwidth]{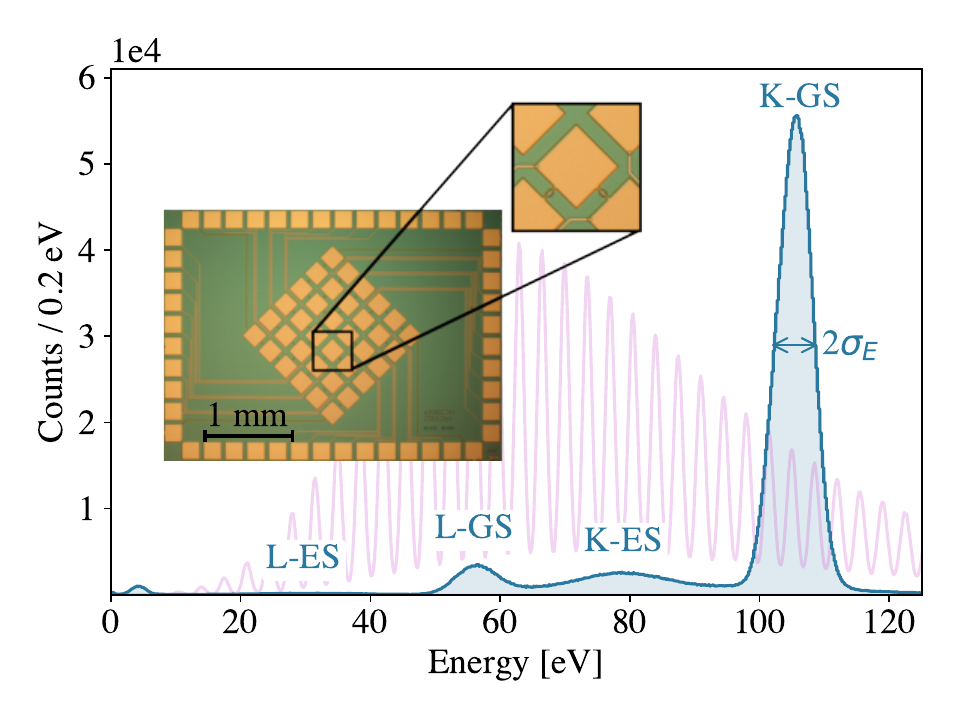}
\caption{\label{fig:spectrum}The measured $^7$Li recoil spectrum with the four peaks described in the text for 20 hours of data from the single STJ pixel shown in the inset.  The measured $\sigma_{\textrm{N},E}$ of the K-GS peak is shown, and is conservatively extracted as the upper limit on the energy width of the recoil.  The calibration laser spectrum (violet) is also shown for comparison.}
\end{figure}

Data were acquired for $\sim$20~hours on November 6, 2022. Voltage traces from the STJ array were continuously recorded along with the laser and processed using a trapezoidal filter.  The pulsed laser was used for energy calibration and drift correction over the period of data collection.  The laser events were selectively identified in the data stream via their coincidence with a logic pulse that was used to trigger the laser, and the comb of peaks is shown in Fig.~\ref{fig:spectrum}.  The laser signal was fit to a superposition of individual Gaussian functions, each corresponding to an integer multiple of the single photon energy.  The peaks were used to generate a quadratic calibration function which was applied to the corresponding $^7$Be spectrum.   The energy resolution of the laser peaks was $1.91(2)$~eV at 105~eV with residuals of $\sim$0.01~eV in the region of interest \cite{Fri20,bray_data_2023}.

The final $^7$Be decay spectrum shown in Fig.~\ref{fig:spectrum} includes events from a single pixel that were not in coincidence with the laser logic pulse or simultaneous signals in other pixels.  The latter of these two criteria was imposed to remove the large, low-energy background from 478~keV $\gamma$-ray Compton scattering interactions in the 0.5~mm Si substrate below the STJ array which generated signals in many pixels simultaneously.  The resulting spectrum contains four main peaks resulting from the two atomic and two nuclear processes described above. The two recoil peaks from the excited state decay of $^7$Li were Doppler-broadened from $\gamma$-decay in flight~\cite{Fre20,PhysRevLett.126.021803}.

\begin{figure}[t!]
\centering
\includegraphics[width=\textwidth, trim=120 20 100 65, clip]{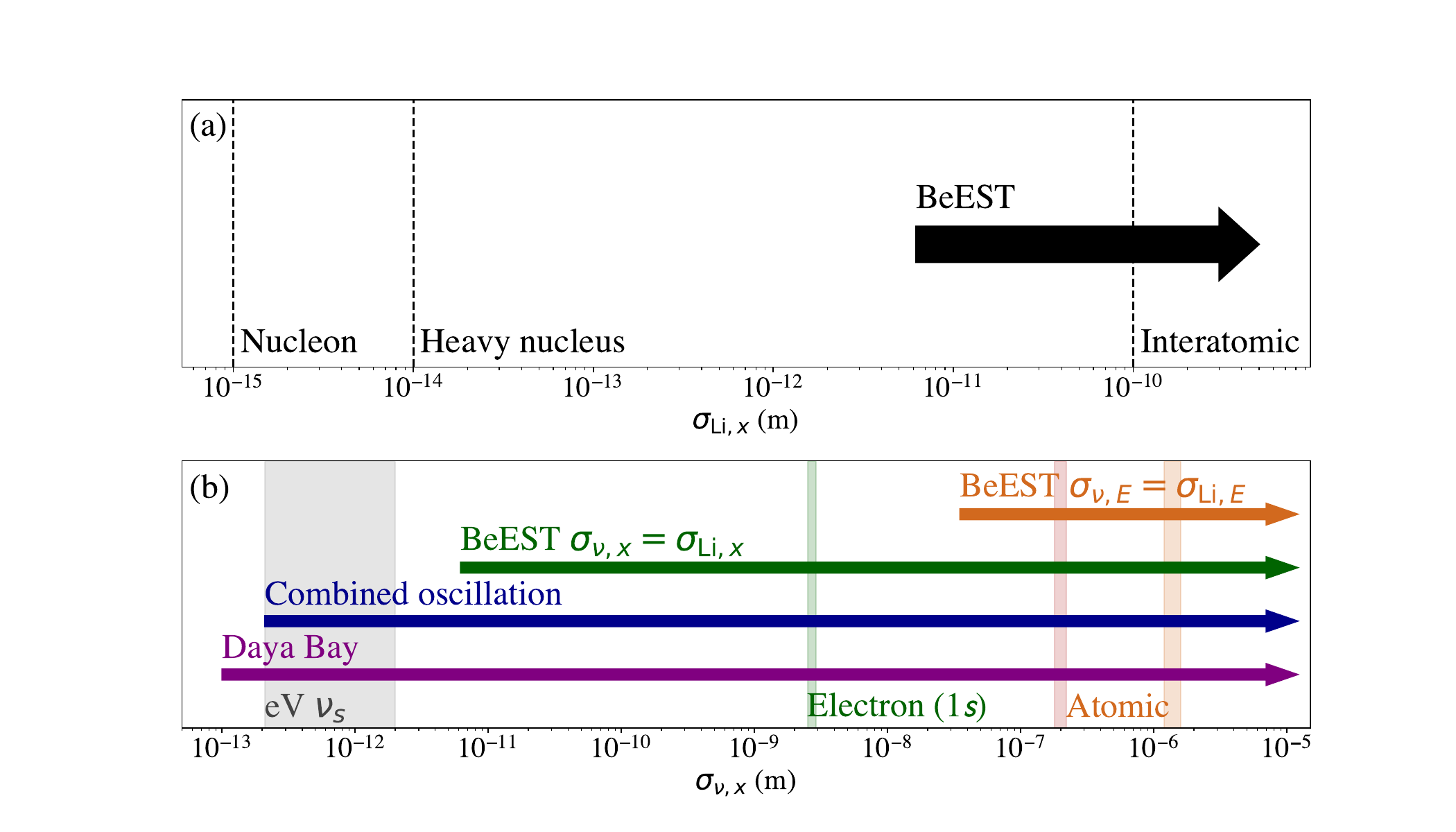}
\caption{\label{fig:limits}(a) The lower limit on the spatial width of the $^7$Li final-state recoil compared to the approximate nuclear and atomic scales.  (b)  BeEST lower limits on spatial width of $\nu_e$ using the two theoretical approaches described in the text compared to limits extracted from reactor oscillation experiments \cite{daya_bay_collaboration_study_2017, de_gouvea_combined_2021}.  The band on the left corresponds to neutrino wavepacket widths that improve eV-scale $\nu_s$ model fits to data \cite{arguelles_impact_2023, hardin_new_2023}, and are excluded by both BeEST limits.  The other bands show predictions based on different methods and localization scales (widened for clarity): (i) The green band around 2.7~nm comes from conservation of momentum with localization on the scale of the $1s$ electron orbital \cite{jones_EC}. (ii) The orange band around 1.4~$\mu$m comes from conservation of energy with localization caused by atomic interactions \cite{akhmedov_damping_2022}. (iii) The red band around 200~nm is based on a QFT approach that also uses atomic interactions for localization \cite{krueger_decoherence_2023}.}
\end{figure}

Since the most probable $^7$Be decay mode is K-shell capture to the nuclear ground state of $^7$Li~\cite{Fre20}, the K-GS peak provides the highest statistical significance for a measurement of the energy width.  The K-GS peak shown in Fig.~\ref{fig:spectrum} was fit in the region from $100-111$\,eV with a single Voigt function \cite{scipy, vugrin_confidence_2007} to extract the energy uncertainty width $\sigma_{\textrm{N}, E} \leq$ 2.9~eV at 95\% confidence level (CL) using the profile likelihood method~\cite{royston_profile_2007}.  Some of this width is due to structural and chemical variations of the Li $1s$ binding energy at different sites in the Ta lattice, which contribute $\mathcal{O}$(1eV) to the broadening~\cite{PhysRevApplied.19.014032}.  However, due to their complexity and the fact that these effects are not yet quantified in detail for the BeEST energy spectrum, we conservatively attributed all of the measured broadening in the K-GS peak to quantum uncertainty.  We then converted the full energy width to a momentum width of the $^7$Li recoil.  The uncertainty was propagated using $\sigma_p = \sqrt{m/2E}\,\sigma_E$ and resulted in $\sigma_{\textrm{Li}, p}\leq 16$~keV/$c$.  Finally, using the Heisenberg uncertainty relation, we extracted a lower limit on the spatial width on the nuclear recoil of $\sigma_{N,x}\geq6.2$~pm, shown in Fig.~\ref{fig:limits} (a).  These results are based on our measurement of the K-GS peak energy width under the experimental conditions reported here and have no underlying assumptions about the scale of localization.

To extract a limit on the neutrino spatial width, $\sigma_{\nu,x}$, two approaches were used: one based on conservation of energy outlined in Ref.~\cite{akhmedov_damping_2022} and the other on conservation of momentum in Ref.~\cite{jones_width_2023}.  In the first method, the neutrino energy width is equivalent to that of the entangled recoil nucleus, which in our case is the upper-limit on the measured energy width $\sigma_{\nu, E} = \sigma_{\textrm{N}, E} \leq 2.9$~eV.  This limit was then converted to a momentum uncertainty using the relativistic energy-momentum relation, $p_\nu = E_\nu / c$ and $\sigma_p = \sigma_E / c$, and the spatial uncertainty was extracted as $\sigma_{\nu,x}\geq35$~nm at 95\% CL using the uncertainty principle.  We also note that this result is the same as a similar approach prescribed in Ref.~\cite{beuthe_oscillations_2003} based on environmental interactions leading to a velocity dependent relation between the spatial width and energy width.

For the second approach reported in Ref.~\cite{jones_width_2023}, the momentum widths of the recoil and nucleus are equivalent, $\sigma_{\nu,p} = \sigma_{\textrm{Li}, p} \leq 16$~keV/$c$, and the uncertainty principle again gives the spatial width.  Under this approach, we determine a lower limit on the spatial width of the neutrino of $\sigma_{\nu,x} = \sigma_{N,x}\geq6.2$~pm at 95\% CL.

Since the two methods described above yield dramatically different results for the neutrino wavepacket width, the details of which are discussed extensively in the literature~\cite{beuthe_oscillations_2003, akhmedov_damping_2022, jones_comment_2022, akhmedov_reply_2022, jones_width_2023, krueger_decoherence_2023, jones_EC}, we do not claim validity of one over the other here, but rather discuss the potential implications of our measurement on the neutrino wavepacket width extracted using both.

First, our extracted limits on the neutrino widths, $\sigma_{\nu,x}\geq6.2$~pm and $\sigma_{\nu,x}\geq35$~nm, are both much larger than the level at which wavepacket separation in eV-scale models could be used to remove tension between reactor experiments and the so-called ``gallium anomaly" in EC decay oscillation experiments~\cite{arguelles_impact_2023, hardin_new_2023}, as shown in Fig.~\ref{fig:limits} (b).  To place our extracted limits within the context of the broader neutrino physics landscape, we then take the standard approach~\cite{arguelles_impact_2023, hardin_new_2023} of assuming that the variations in $\sigma_{\nu, x}$ between types of sources ($\beta^\pm$ or EC) is small compared to variations between scales of localization.  Under this treatment, our lower-limit on the neutrino width extracted using the first approach is more stringent than the previous experimental limits on $\sigma_{\nu,x}$ by reactor sources by more than an order of magnitude.  Additionally, our limits using either method exclude the possibility of future experiments such as JUNO to be able to observe wavepacket separation~\cite{juno_collaboration_damping_2022, de_gouvea_probing_2020}.  These results have implications beyond what is discussed here, however for clarity and model-independence of what we report we only present the above discussion.

Beyond the implications for neutrino physics, perhaps the more important result from our work is the potential to use nuclear EC decay as a fundamental test of quantum mechanics at subatomic scales.  Since the electron is captured and the neutrino is produced within the nuclear volume via weak coupling at the quark level, there is a question as to whether processes such as short-range correlations~\cite{RevModPhys.89.045002} are able to decohere and localize the system via interactions within the atomic nucleus ($<$10~fm).  Our experimental determination of the position width of the daughter recoil, $\sigma_{\textrm{N}, x} \geq 6.2$\,pm, excludes this possibility by several orders of magnitude.  This general conclusion is consistent with nuclear $\gamma$ decay, an electromagnetic process, in highly localized environments through the M\"ossbauer effect, which suggests localization on the scale of 1-10~pm~\cite{Mossbauer}.

In summary, we present a new experimental concept using highly localized EC decaying rare isotopes inside superconducting sensors to measure the quantum properties of the final-state system in electroweak nuclear decay.  Here, we report an upper limit of the energy width of the recoiling $^7$Li system from the EC decay of $^7$Be in a tantalum matrix at 0.1~K of $\sigma_{E,N}\leq2.9$~eV (95\% CL).  This limit corresponds to a limit on the spatial width of the $^7$Li nuclear recoil of $\sigma_{\textrm{N}, x} \geq 6.2$\,pm, which suggests that localization of the final state system occurs at a scale much larger than the nucleus in which the interaction occurs.  From this measurement, we extract limits on the spatial width of the entangled neutrino using two methods reported in the literature, the first such limits on an EC $\nu_e$ source to date.  Although the extraction of this value depends dramatically on the theoretical prescriptions used, both methods give neutrino wavepacket sizes that are too large to observe wavepacket separation in future large scale oscillation experiments such as JUNO, and does not relieve the tension in 3+1 $\nu_s$ model fits. These results demonstrate the power of using precision measurements of EC decay as a probe of fundamental physics, and the reported results have implications for the details of neutrino oscillation and BSM theories, as well as serving as a test of quantum mechanics at the subatomic scale.

\section*{Acknowledgements}
We thank David Moore, Daniel Carney, Benjamin Jones, Carlos Arg\"uelles-Delgado, Joseph Formaggio, and Fred Sarazin for useful discussions.  The BeEST experiment is funded by the Gordon and Betty Moore Foundation (10.37807/GBMF11571), the U.S. Department of Energy Office of Science, Office of Nuclear Physics under Award Numbers DE-SC0021245 and SCW1758, the LLNL Laboratory Directed Research and Development program through Grants No. 19-FS-027 and No. 20-LW-006, the European Metrology Programme for Innovation and Research (EMPIR) Projects No. 17FUN02 MetroMMC and No. 20FUN09 PrimA-LTD, and the FCT—Fundação para a Ciência e Tecnologia (Portugal) through national funds in the framework of the projects UID/04559/2020 (LIBPhys). TRIUMF receives federal funding via a contribution agreement with the National Research Council of Canada. This work was performed under the auspices of the U.S. Department of Energy by Lawrence Livermore National Laboratory under Contract No. DE-AC52-07NA27344.  Francisco Ponce is supported by Pacific Northwest National Laboratory, which is managed for the US Department of Energy by Battelle under contract DE-AC05-76RL01830.  KGL gratefully acknowledges support from the Facility for Rare Isotope Beams (FRIB) while on sabbatical.  FRIB is a US Department of Energy, Office of Science User Facility under Award Number DE-SC0000661.

\bibliography{sn-bibliography}

\end{document}